\documentclass[twocolumn,tighten,times]{aastex63}

\usepackage{amsmath,amsfonts,amssymb, mathtools,multirow}

\accepted{March 13, 2025}

\submitjournal{AJ}

\shorttitle{Cluster Galaxy Evolution}
\shortauthors{Lin et al.}

\begin{document}

\title{Evolution of Massive Red Galaxies in Clusters from $z=1.0$ to $z=0.3$}

\author[0000-0001-7146-4687]{Yen-Ting Lin}
\affiliation{Institute of Astronomy and Astrophysics, Academia Sinica, Taipei 10617, Taiwan}
\affiliation{Institute of Physics, National Yang Ming Chiao Tung University, Hsinchu 30010,  Taiwan}
\affiliation{Institute for Advanced Study, Princeton, NJ 08540, USA}
\affiliation{Department of Astrophysical Sciences, Princeton University, Princeton, NJ 08542, USA}

\author[0000-0002-3839-0230]{Kai-Feng Chen}
\affiliation{Department of Physics, Massachusetts Institute of Technology, Cambridge, MA 02139, USA}

\author[0009-0008-7522-4179]{Tsung-Chi Chen}
\affiliation{Department of Physics, National Taiwan University, Taipei 10617, Taiwan}
\affiliation{Institute of Astronomy and Astrophysics, Academia Sinica, Taipei 10617, Taiwan}

\author[0000-0003-2069-9413]{Chen-Yu Chuang}
\affiliation{Institute of Astronomy and Astrophysics, Academia Sinica,   Taipei 10617, Taiwan}
\affiliation{Department of Astronomy, University of Arizona, Tucson, AZ 85721, USA}

\author[0000-0003-3484-399X]{Masamune Oguri}
\affiliation{Center for Frontier Science, Chiba University,  Chiba 263-8522, Japan}



\begin{abstract}

A critical issue in studying the evolution of galaxy clusters is to find ways that enable meaningful comparisons of clusters observed at different redshifts, as well as in various stages of their growth.
Studies in the past have typically suffered from uncertainties in cluster mass estimates due to the scatter between cluster observables and mass.
Here we propose a novel and general approach that uses the probability distribution function of an observable--cluster mass relation, together with dark matter halo merger trees extracted from numerical simulations, such that one can trace the  evolution in a self-contained fashion, for clusters chosen to lie in a specified  range in mass and redshift.  This method, when applied to clusters of different mass ranges, further allows one to examine the evolution of various observable--cluster mass scaling relations.  
We illustrate the potential of this method by studying the stellar mass content of red cluster member galaxies, as well as the growth of brightest cluster galaxies, from $z=1.0$ to $z=0.3$, using a large optically-detected cluster sample from the Subaru Hyper Suprime-Cam Survey, finding good agreement with previous studies.

\end{abstract}



\section{Introduction} 
\label{sec:intro}

A central question concerning galaxy evolution is how to fairly compare galaxy populations across a wide range in cosmic time.  Ideally, one wishes to compare low redshift galaxies with those that could represent their progenitors.
For field galaxies, it was suggested that the cumulative number density selection method can produce such a progenitor--descendant relationship, thus allowing one to trace the growth of galaxies over time \citep[e.g.,][]{vandokkum10,leja13}.

For galaxies in clusters, in principle, to achieve such a goal is easier, as the formation history of massive dark matter halos is less stochastic than that of lower mass halos that host field galaxies (e.g., \citealt{lacey93,zhao09}).  
Several studies have made use of dark matter halo growth histories (also known as merger trees) to make the connection between high-$z$ and low-$z$ clusters (e.g., \citealt{lidman12,lin13,vanderburg15,shankar15}).

A ``Top-$N$'' selection method, similar in spirit to the cumulative number density selection mentioned above, was proposed and tested against numerical simulations \citep{inagaki15,lin17}.  The method relies on the Ansatz that, given a comoving volume, the most massive $N$ halos at one epoch shall remain among the most massive $N$ in another epoch that is not too distant in time.  Using the Millennium Run \citep{springel05}, it was found that the Ansatz is only partially true -- at least 66\% of the most massive 100 halos over the $(500\,h^{-1}\,{\rm Mpc})^3$ volume would remain among the most massive, irrespective of the redshift difference (can be as large as $\Delta z=0.6$, or over a time interval of $\sim 3$\,Gyr).  Nevertheless, one can still faithfully recover the evolution of the luminosity function and stellar mass function of cluster galaxy population with such an approach \citep{lin17}.

Using the cluster sample constructed from the first-year data of the Hyper Suprime-Cam Subaru Strategic Program (HSC-SSP; \citealt{aihara18,miyazaki18b}), \citet{lin17} studied the cluster galaxy population evolution from $z\approx 1$ to $z=0.3$, for the top 100 {\it richest} clusters in 4 redshift bins detected by the {\sc camira} algorithm \citep{oguri14,oguri18},
over an area of 230\,deg$^2$.  Based on  both the abundance of clusters and weak lensing mass measurements, they suggested that the selected clusters follow the evolutionary track of clusters whose present-day mass is $M_{200c} \sim 8\times 10^{14}\,h^{-1}M_\odot$.\footnote{$M_{200x}$ (where $x=c$  or $m$) is defined as the total mass within a radius $r_{200x}$, within which the mean density is 200 times of the critical ($x=c$) or mean ($x=m$) density of the Universe at the redshift of the cluster.  Typically $M_{200m}$ is about $20-25\%$ larger than $M_{200c}$.}

\begin{figure*}[]
    \epsscale{0.8}
    \plotone{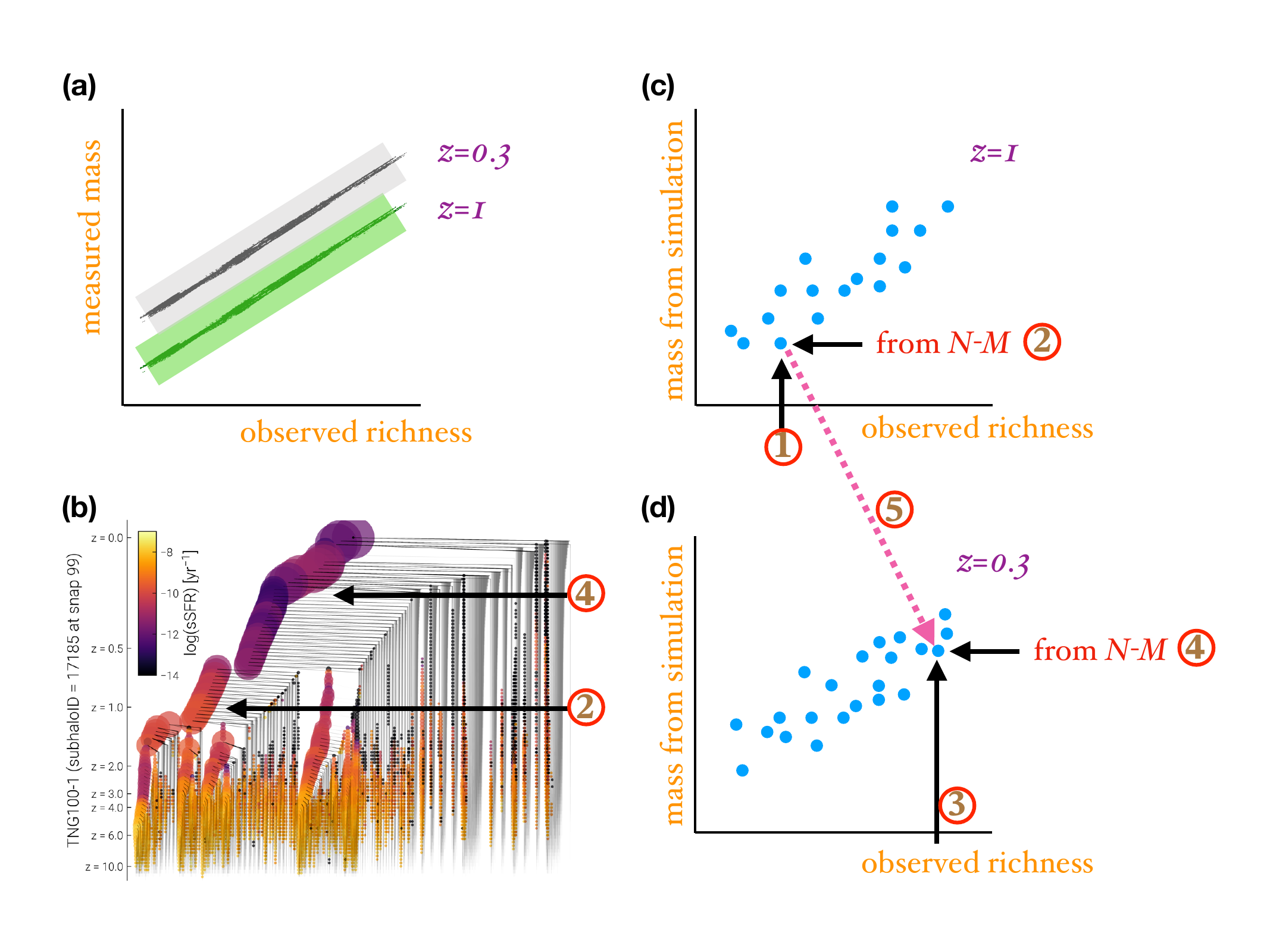}
    \caption{ Illustration of our method.  Panels (a) is a sketch of the mass--richness  ($N$--$M$) relation measured at two different redshifts,  at e.g., $z=1$ (green) and $z=0.3$ (black); the line and shaded regions represent relations and the associated uncertainties).  Panel (b) is an example merger tree; the size and the color of the circles represent the mass and specific star formation rate of the halos.  For our method, {\it merger trees from a dark matter-only simulation are sufficient.}
    We shall use panels (b), (c) and (d) to demonstrate our approach.
    Briefly speaking, one starts from a real cluster at high-$z$ (step 1 in panel c); using the measured $N$--$M$ relation at that redshift (e.g., green line in panel a), one infers a corresponding halo mass (step 2 in panel c), and try to find the closest halo in terms of mass at that redshift (step 2 in panel b).  For a real cluster at low-$z$ (step 3 in panel d), we use the low-$z$ $N$--$M$ relation to find the most appropriate mass (step 4 in panel d), then search for the best matched halo at that redshift (step 4 in panel b).  This way, one can ``connect'' two observed clusters at the two cosmic epochs and study how the galaxy populations evolve (step 5). 
    Panel (b) is generated courtesy of the IllustrisTNG collaboration \citep{nelson19}.  
    }
    \label{fig:sum} 
\end{figure*}

In principle, by changing $N$ from 100 to larger numbers, one can study the evolution of clusters whose present-day mass is lower.  However, one limitation of this method is that one can only select clusters by the observable, in this case the richness (throughout this paper, defined as the number of {\it red} member galaxies more massive than $\approx 10^{10}h^{-1} M_\odot$; \citealt{oguri18}), and there is a considerable scatter between it and the true halo mass (e.g., \citealt{simet17,murata19}).  It is also impractical to use this method to study the evolution of clusters within a particular mass range, should one desire to do so.

We have developed
a new method that makes heavy use of dark matter halo merger trees (so that the progenitor--descendant relationship for clusters is intrinsically built-in), while simultaneously taking into account the scatter in the richness--cluster mass relation ($N$--$M$ for short).
With this  approach, it becomes feasible to trace the evolution of clusters of a given mass at the present-day (or some chosen redshift).  By performing the method for clusters of several mass bins, it is also possible to study the redshift evolution of scaling relations between some observable and cluster mass.

In this paper, which  focuses on the ideas behind the methodology, we illustrate the approach by studying the evolution of the stellar mass content of {\it red} member galaxies of {\sc camira} clusters from $z=1.0$ to $z=0.30$.  
The paper is structured as follows: in Section~\ref{sec:method} the basic idea and the implementation of the method are described.  In Section~\ref{sec:res} we present our main result, namely the evolution of the red galaxy population, as well as the mass growth history of brightest cluster galaxies (BCGs).  We discuss the potential applications of this method in Section~\ref{sec:disc}.
Throughout the paper we adopt a WMAP7 \citep{komatsu11} $\Lambda$CDM model, with $\Omega_m=0.272$, $\Omega_\Lambda=0.728$, $H_0=100 h$\,km\,s$^{-1}$\,Mpc$^{-1}$ with $h=0.704$, and $\sigma_8=0.81$.

\begin{table*}[ht]
\caption{Fitting Results of Scaling Relations}
\begin{tabular}{ cccccccc } 
 \hline
\multicolumn{1}{c}{Bin} & \multicolumn{1}{c}{Redshift} & \multicolumn{1}{c}{Mean $z$} & \multicolumn{1}{c}{MR } & \multicolumn{1}{c}{number of} &  \multicolumn{1}{c}{$\alpha$: slope of} & \multicolumn{1}{c}{$\beta$: slope of}  & \multicolumn{1}{c}{$\gamma$: slope of} \\

\multicolumn{1}{c}{} & \multicolumn{1}{c}{range} & \multicolumn{1}{c}{} & \multicolumn{1}{c}{ snapshot} & \multicolumn{1}{c}{clusters} & \multicolumn{1}{c}{$M_{\rm star,tot}$--$M_{200m}$}  & \multicolumn{1}{c}{$M_{\rm bcg}$--$M_{200m}$}  & \multicolumn{1}{c}{$N$--$M_{200m}$} 
\\
 \hline
1 & $0.30-0.55$ & $0.42$ & 47 & 3483 & $0.68 \pm 0.08$ & $0.24 \pm 0.13$ & $0.82\pm 0.07$\\
2 & $0.55-0.69$ & $0.62$ & 44 & 3398 & $0.68 \pm 0.08$ & $0.26\pm 0.13$ & $0.80 \pm 0.06$\\
3 & $0.69-0.81$ & $0.75$ & 42 & 2640 & $0.64 \pm 0.09$ & $0.30\pm 0.12$ & $0.77 \pm 0.06$\\
4 & $0.81-0.90$ & $0.86$ & 41 & 2372 & $0.70 \pm 0.08$ & $0.28 \pm 0.15$ & $0.82 \pm 0.05$\\
5 & $0.90-0.99$ & $0.95$ & 40 & 2331 & $0.67 \pm 0.09$ & $0.39 \pm 0.15$ & $0.75 \pm 0.06$\\

 \hline
  \label{tab:res}
\end{tabular}
\end{table*}

\section{Method} 
\label{sec:method}

\subsection{Overview}
\label{sec:over}

The basic idea of our method is 
in spirit similar to halo abundance matching (e.g., \citealt{conroy06,trujillo-gomez11}), and is
summarized in Figure~\ref{fig:sum}.  Shown on panel (a) is an illustration of measured relation between cluster mass $M_{200m}$ and richness $N$ at two redshifts (black: high-$z$; green: low-$z$). The shaded region represents the non-negligible scatter of the relation (see Section~\ref{sec:nm}).  Panel (b) is an example of merger trees; in our approach, many such trees for cluster-scale halos will be utilized (which in turn requires a large $N$-body simulation box).  We proceed as follows: in step 1, at a certain high redshift bin (with a mean redshift  $z_1=1$),  for each cluster belong to that bin, from its observable, namely richness $N$, we infer the most likely cluster mass from the $N$--$M$ relation (steps 1 \& 2 in panel c).  We then go through all branches of all merger trees at that $z_1$ from a simulation, finding the halo with a mass that is closest to the cluster mass (step 2 in panel b).
At a lower redshift bin (with a mean redshift $z_2=0.3$), we repeat the exercise (steps 3 \& 4 in panel d), and find a branch 
with closest halo mass as predicted from the $N$-$M$ relation at $z_2$.
When all clusters in all redshift bins are assigned to a halo at appropriate snapshots of the merger trees, we can then, for each branch at $z_1$, find all the descendant halos and the clusters associated with them, and compute the mean properties of the clusters, which enables us to statistically connect clusters at different cosmic epochs together (step 5: connecting clusters from panels c and d).  Given the scatter of the $N$--$M$ relation, we need to repeat the whole procedure many times to get the averaged behavior.
For the actual implementation of the method, please see Section~\ref{sec:actual}.

In practice, we have chosen 5 redshift intervals, each occupying the same comoving volume, between $z=0.300$ and $z=0.992$.  
At $z<0.3$, there is simply not enough survey volume.  Furthermore, the {\sc camira} sample is available at $z\ge 0.1$, further reducing the available volume.  This is the primary reason we set $z=0.3$ as our lower limit in redshift.
Our upper limit in redshift is set to $z\approx 1$ because 
 (1) the $N$--$M$ relation we will emplloy was only measured up to that redshift, and (2) no incompleteness in richness was applied to {\sc camira} clusters at $z>1$.
In the first three columns of Table~\ref{tab:res}, we show some basic information of the redshift bins.
In the following, we provide  details of the ingredients needed for our method, namely the cluster sample (Section~\ref{sec:cls}), the merger trees (Section~\ref{sec:milsim}), and the $N$--$M$ relation and the associated probability distribution function (PDF) $P(M|N,z)$ 
measured at $z=0.1-1$
(Section~\ref{sec:nm}).

\begin{table*}[ht]
\centering
\caption{Parameters of the mass--richness relation used in this work}
\begin{tabular}{ clc } 
 \hline
\multicolumn{1}{c}{Parameter} & \multicolumn{1}{c}{Description} & \multicolumn{1}{c}{Median} \\
 \hline
 $A$ & $\langle\ln N\rangle$ at pivot mass scale and pivot redshift & $3.36$ \\ 
 $B$ & Coefficient of halo mass dependence in $\langle\ln N\rangle$ & $0.83$ \\ 
 $B_z$ & Coefficient of linear redshift dependence in $\langle\ln N\rangle$ & $-0.20$ \\ 
 $C_z$ & Coefficient of square redshift dependence in $\langle\ln N\rangle$ & $3.51$ \\
 $\sigma_0$ & $\sigma_{\ln N \mid M, z}$ at pivot mass scale and pivot redshift & $0.19$ \\
 $q$ & Coefficient of halo mass dependence in $\sigma_{\ln N \mid M, z}$ & $-0.02$ \\
 $q_z$ & Coefficient of linear redshift dependence in $\sigma_{\ln N \mid M, z}$ & $0.23$ \\
 $p_z$ & Coefficient of square redshift dependence in $\sigma_{\ln N \mid M, z}$ & $1.26$ \\
 \hline
  \label{tab:parameter}
\end{tabular}
\end{table*}

\subsection{Cluster Sample}
\label{sec:cls}

We use the latest public version of the HSC {\sc camira} sample (version S20A), which is based on an area of 780\,deg$^2$.  
{\sc camira} is a multi-band red-sequence cluster finding algorithm that incorporates filters in member galaxy spatial distribution \citep{oguri14}.
Above richness $N=10$, there are 14,224 clusters between $z=0.30$ and $1.0$ (see the fifth column of Table~\ref{tab:res}).  For each cluster, {\sc camira} provides an estimate of the total stellar mass content of red member galaxies (with a lower limit of $\approx 10^{10}M_\odot$; \citealt{oguri18}), $M_{\rm star,tot}$, which takes into account the membership probability.  The photometry used to compute the total luminosity of individual galaxies is the {\tt cmodel} magnitude \citep{bosch18}.  The most massive galaxy of a cluster, with a probability of being member exceeding 0.7, is chosen as the BCG.

\subsection{Merger Trees}
\label{sec:milsim}

We extract the merger trees from the version of the Millennium Run (MR) simulation  re-run with the WMAP7 cosmology.  Specifically, we query the Millennium database to retrieve the merger trees of all halos that satisfy $M_{200c}\ge 7\times 10^{13} h^{-1}\,M_\odot$ at $z=0.32$ and $M_{200c}\ge 3\times 10^{12} h^{-1}\,M_\odot$ at $z=1.28$.\footnote{The reason our query uses $M_{200c}$ is because this mass is indexed in the database; querying using $M_{200m}$ always results in timeout.  Our lower limit in mass of the progenitors is sufficiently low that our way of query does not affect our results. In addition, the choice of limiting mass of $M_{200c}=3\times 10^{12}h^{-2}M_\odot$ is to ensure an $N=10$ cluster at $z=1.28$ can have a matched counterpart (see more details in Section~\ref{sec:actual}), but in the actual analysis we limit ourselves to clusters and halos at $z\le 1$.}  
Our 5 redshift bins correspond to snapshots 47, 44, 42, 41, and 40 (from low to high redshift; see the 4th column of Table~\ref{tab:res}).  
In some rare cases, when a branch (or even the main trunk) of a merger tree misses data 
between some snapshots, 
for sake of simplicity, we ignore it from the analysis.

Furthermore, given that the volume of MR $(500\,h^{-1}\,{\rm Mpc})^3$ is much smaller than that probed by the HSC survey (out to $z=1$), we have chosen our redshift bins such that the comoving volume in each bin is somewhat smaller to that of MR, i.e., $(476\,h^{-1}{\rm Mpc})^3$, so that there are more halos (equivalent to branches of all merger trees at a given snapshot) than clusters in each redshift bin.  
However, such a choice corresponds to a sky area of 390\,deg$^2$, which is half of that covered by the HSC survey.
We therefore split the {\sc camira} cluster sample into two halves (hereafter referred to as sets I and II), so that we have comparable numbers of merger trees and clusters (although there are always more halos than clusters, particularly at high-$z$).  Going through the {\sc camira} cluster catalog, which is ordered in increasing order of right ascension, we simply group all the odd-numbered clusters into set I, and even-numbered ones into set II, which results in nearly identical distributions in redshift and richness for the two sets of clusters.\footnote{We have also tried splitting clusters into two sub-samples with roughly the same number of clusters by a simple division in terms of right ascension, and the results as shown in Section~\ref{sec:res} remain unchanged.}

\subsection{The Richness--Cluster Mass Scaling Relation}
\label{sec:nm}

Through a combined analysis of stacked lensing profiles and cluster abundance, \citet{murata19} constrained the $N$--$M$  relation of the  {\sc camira} cluster sample assuming a log-normal distribution
\begin{equation}
P(\ln N \mid M, z)=\frac{1}{\sqrt{2 \pi} \sigma_{\ln N \mid M, z}} \exp \left[-\frac{x^{2}(N, M, z)}{2 \sigma_{\ln N \mid M, z}^{2}}\right],
\end{equation}
where $x(N, M, z)$ models the mean mass--richness relation with the functional form
\begin{equation}
\begin{aligned}
x(N, M, z) \equiv & \ln N-\left\{A+B \ln \left(\frac{M}{M_{\text {pivot }}}\right)\right.\\
&+B_{z} \ln \left(\frac{1+z}{1+z_{\text {pivot }}}\right) \\
&\left.+C_{z}\left[\ln \left(\frac{1+z}{1+z_{\text {pivot }}}\right)\right]^{2}\right\} .
\end{aligned}
\end{equation}
and the scatter $\sigma_{\ln N \mid M, z}$ is parametrized as
\begin{equation}
\begin{aligned}
\sigma_{\ln N \mid M, z}=& \sigma_{0}+q \ln \left(\frac{M}{M_{\text {pivot }}}\right)+q_{z} \ln \left(\frac{1+z}{1+z_{\text {pivot }}}\right) \\
&+p_{z}\left[\ln \left(\frac{1+z}{1+z_{\text {pivot }}}\right)\right]^{2}
\end{aligned}
\end{equation}
where $M_{\text {pivot} }= 3\times10^{14}h^{-1}M_\odot$ and $z_{\text {pivot} } = 0.6$. 
The parameters  used in these equations are summarized in Table \ref{tab:parameter}.  
For notational simplicity, $M_{200m}$ is  denoted as $M$ throughout this Section.

\begin{figure*}
    \centering  
    \epsscale{0.8}
    \plotone{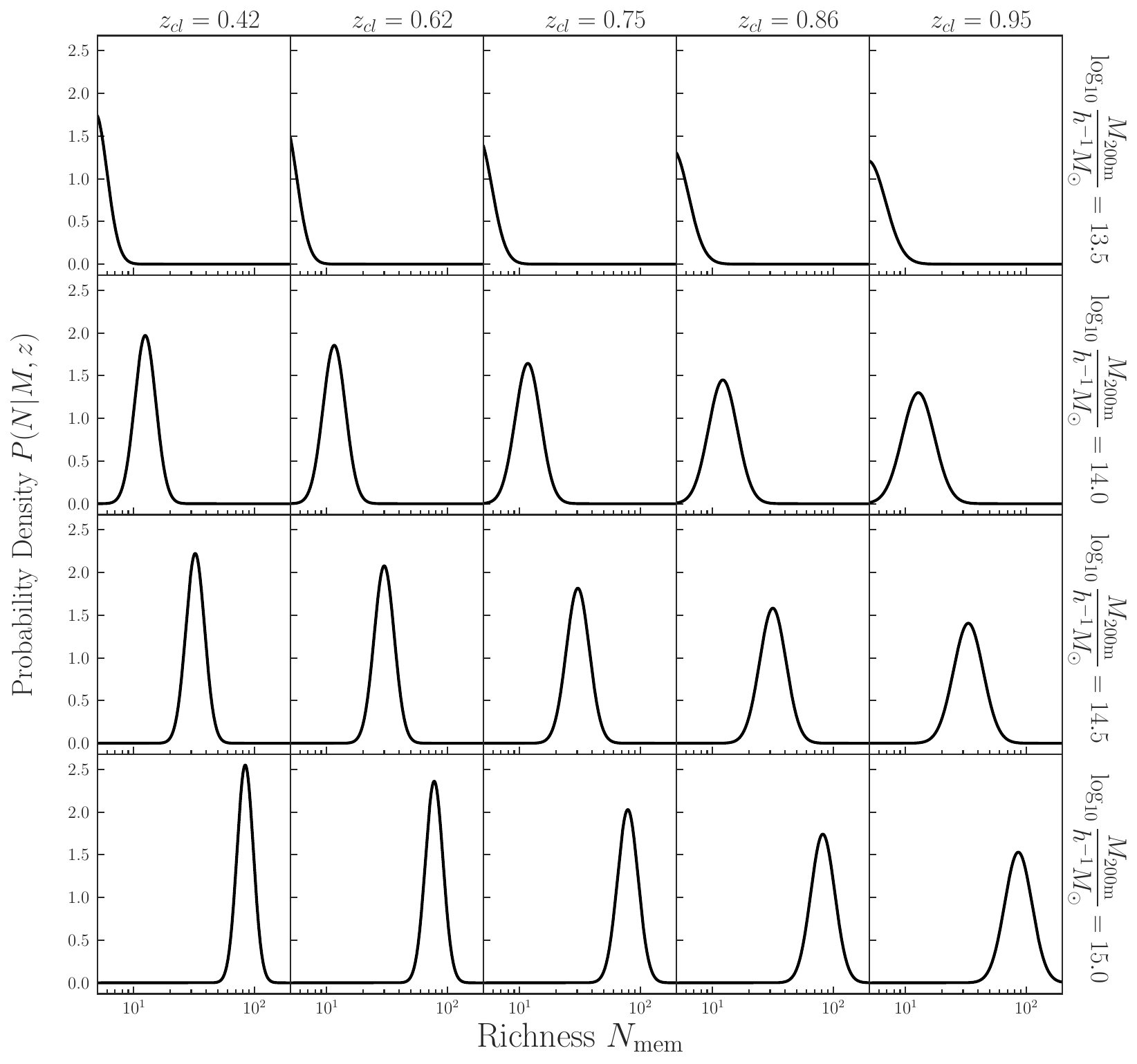}
    \caption{Probability density of richness for different mass and redshift ranges. \label{fig:PofN} }
\end{figure*}

\begin{figure*}
    \centering  
    \epsscale{0.8}
    \plotone{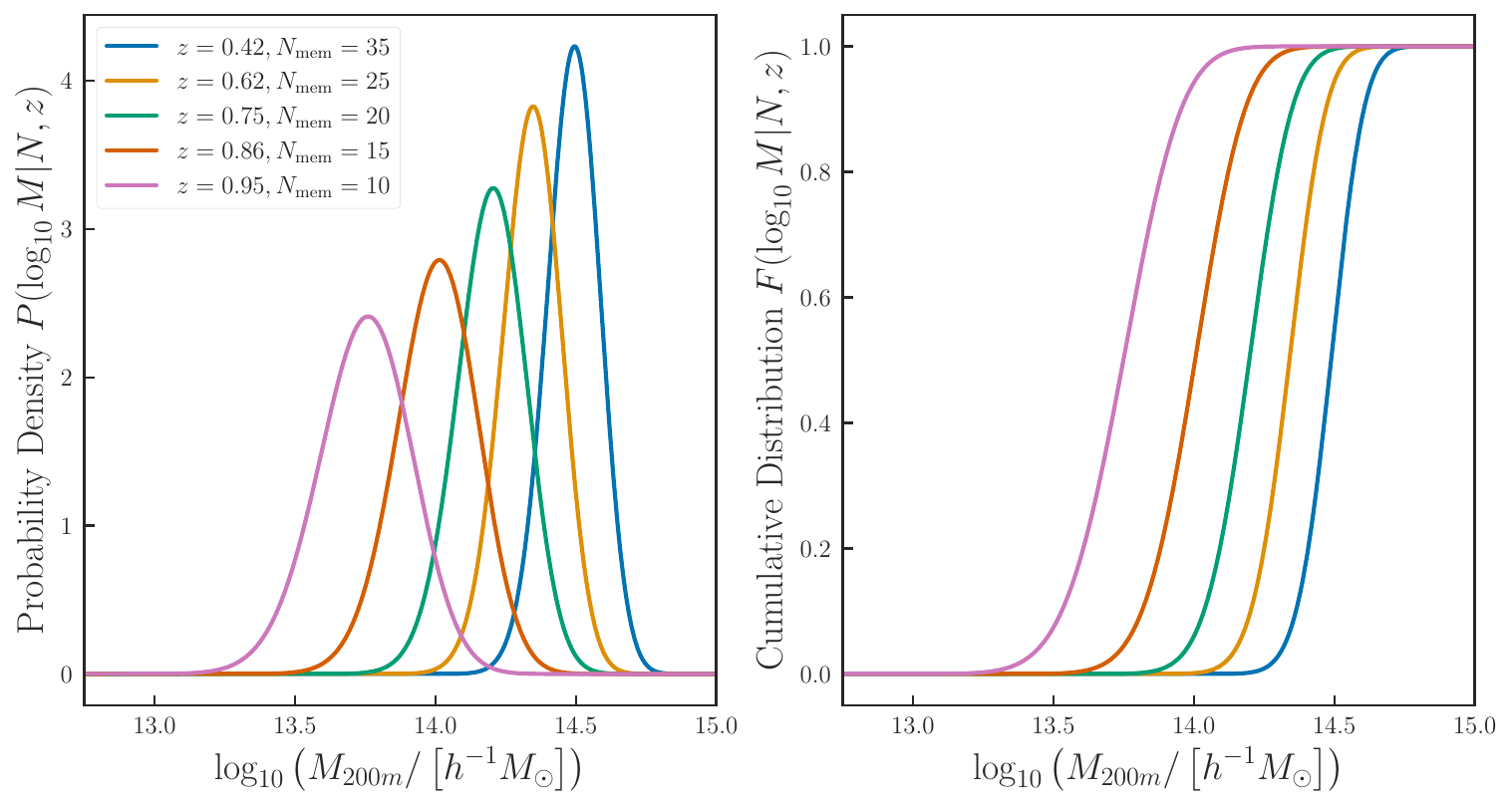}
    \caption{Probability density and cumulative distribution of $\log_{10}(M_{200m})$ for various richness and redshift. \label{fig:PofM} }
\end{figure*}

The mass distribution of a cluster given its richness and redshift can then be obtained through the Bayes' theorem
\begin{equation}
    P(M | \ln N, z) \propto P(\ln N | M, z)\times P(M | z)
\end{equation}
where $P(M | z)$ is further proportional to the halo mass function $\mathrm{d}n/\mathrm{d}M$ at a given redshift. The overall normalization constant $\mathcal{N}$ is thus given by
\begin{equation}
    \displaystyle\mathcal{N}\coloneqq \int P(\ln N | M, z) \frac{\mathrm{d}n(M, z)}{\mathrm{d}M} \mathrm{d}M.
\end{equation}
Here, we adopt the halo mass function from \citet{tinker08}. Therefore, 
\begin{equation}
    P(M | \ln N, z) = \frac{1}{\mathcal{N}}\frac{\mathrm{d}n(M, z)}{\mathrm{d}M}P(\ln N | M, z),
\end{equation}
or more conveniently, 
\begin{equation}
    P(\log_{10}M | \ln N, z) = \frac{1}{\mathcal{N}}\frac{\mathrm{d}n}{\mathrm{d}\log_{10} M}P(\ln N | M, z).
\end{equation}

For the  computation of $P(\log_{10} M|\ln N,z)$, we adopt the constraints under the WMAP9 cosmological model\footnote{\citet{murata19} considered only two sets of cosmological models: WMAP9 and Planck2015.} \citep[][$\Omega_m=0.279$, $\Omega_\Lambda=0.721$, $h=0.70$, $\sigma_8=0.82$]{hinshaw13}, which is close to that of WMAP7. 
Examples of $P(N|M, z)$ for a set of halo masses and redshifts are shown in Figure~\ref{fig:PofN}.
Figure~\ref{fig:PofM} shows the probability density and the cumulative distribution for the inverted scaling relations.

\subsection{Cluster--Halo Assignment}
\label{sec:actual}

With the PDF $P(\log_{10} M_{200m}|\ln N,z)$ in hand, for each cluster, knowing its $N$, we can draw a mass in a Monte Carlo fashion.  Thus, in practice, for step 2 outlined in Section~\ref{sec:over}, we actually choose the cluster whose mass as inferred from its PDF is closest to the halo mass of the merger tree.

We start from the highest redshift bin, and consider all the branches of all merger trees; for each branch, we go through the list of clusters (sorted from richest to poorest) to find one that is closest to its mass and assigned the richness $N$, total  stellar mass $M_{\rm star, tot}$, and BCG stellar mass $M_{\rm bcg}$  to that branch.  The cluster is then removed from the list, and we move to the next branch/halo.
Given that there are more halos than clusters in each redshift bin, all clusters are always assigned a halo.
This procedure is done separately for cluster sets I and II.\footnote{This means that the same branch may be associated with one cluster from each of the cluster sets.  If we only use one cluster set, the results are very similar to those obtained by combining two cluster sets.  See also the comments at the end of Section~\ref{sec:evsc}.}
After going through all the branches in the redshift bin, we compute the averaged $N$,  $M_{\rm star, tot}$ and $M_{\rm bcg}$, separately for halos that would have $\log M_{200m}/(h^{-1}M_\odot)=14.2-14.4$, $14.4-14.6$, and $\ge 14.6$ at $z=0.3$ (hereafter the low, middle, and high mass bins, respectively).
As the cluster--halo assignment based on the mass drawn in the Monte Carlo way is not unique, we repeat the whole procedure 10 times, and use the mean and scatter of the 20 realizations (combining results from sets I and II) as the final results.  Our results do not depend strongly on the actual number of realizations (i.e., using 10, 20, or 40 realizations yields similar outcomes).

The reason we choose the lower limit of the low mass bin to be $\log M_{200m}/(h^{-1}M_\odot)=14.2$ is to avoid incompleteness in our highest redshift bin; 
we have checked that the lowest mass merger tree branch used in our analysis in the highest redshift bin has a mass of $\log M_{200m}/(h^{-1}M_\odot)=13.3$.  From Figure~\ref{fig:PofM} (pink curve on the right panel), we see that only $0.5\%$ of the time a $N=10$ cluster at that redshift would have mass lower than that value, so it will not affect our analysis.

\section{Results} 
\label{sec:res}

\subsection{Evolution of Red Galaxy Content}
\label{sec:evol}

\begin{figure}
    \plotone{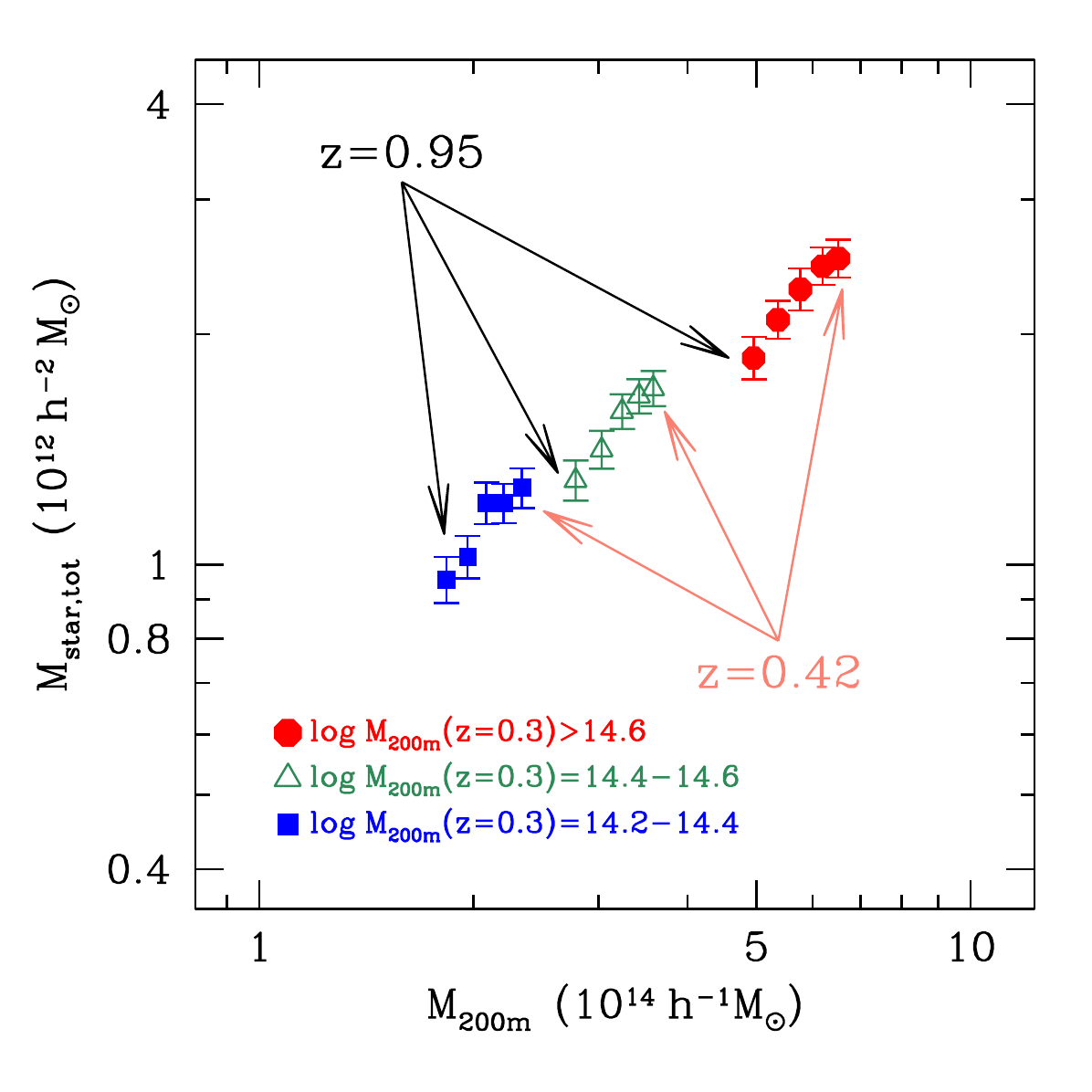}
    \caption{ Redshift evolution of total stellar mass contained in red member galaxies, for three sets of clusters whose $z=0.3$ masses lie in the ranges $\log M_{200m}/(h^{-1}\,M_\odot)=14.2-14.4$ (blue squares), $14.4-14.6$ (green triangles), and $\ge 14.6$ (red solid points).  For each set, the points from lower left to upper right show how $M_{\rm \star,tot}$ evolves with redshift.  In other words, the $x$-axis not only shows the cluster mass, but can also be regarded as the arrow of time (for a given cluster set).  The offset of 3 sets of data points reflects the true halo masses derived from the merger trees (i.e., {\it not} artificially adjusted for clarity of presentation).
     }
    \label{fig:newm} 
\end{figure}

\begin{figure}
    \plotone{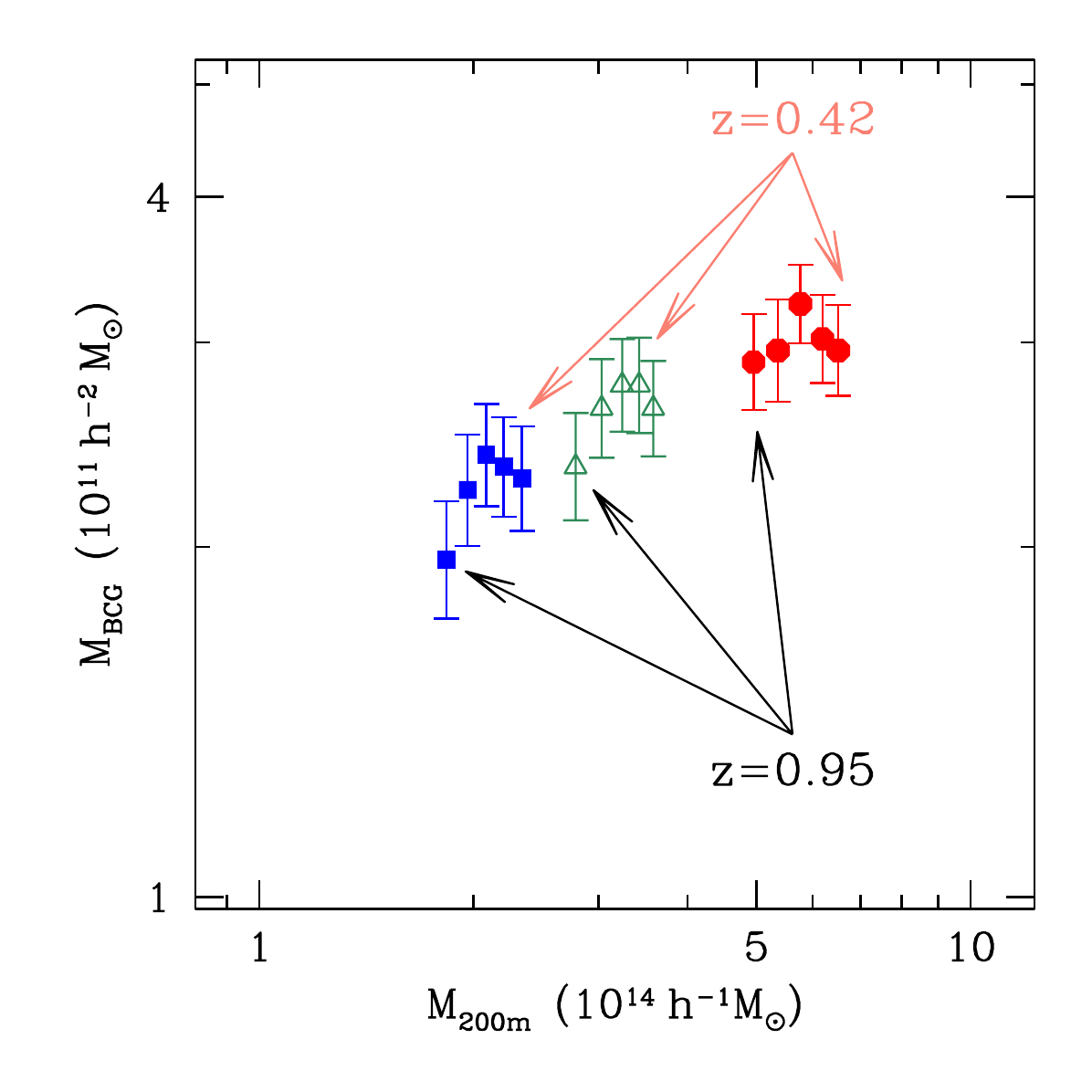}
    \caption{ Similar to Figure~\ref{fig:newm}, but for the stellar mass assembly history of BCGs.  There appears to be a halt in BCG stellar mass growth at $z\lesssim 0.6$.  It is also seen that the BCGs in the high cluster mass bin do not experience much growth from $z=0.99$ to $z=0.3$.
     }
    \label{fig:newb} 
\end{figure}

In Figure~\ref{fig:newm} we show the evolution of the stellar mass content of red cluster galaxies, for clusters 
in the low (blue squares), middle (green triangles), and high (red solid points) mass bins.
Take the high mass bin for example, the mean $M_{\rm star,tot}$ is $(1.9\pm 0.1) \times 10^{12}\,h^{-2}\,M_\odot$ at $z\approx 0.95$ (the mean redshift of our highest redshift bin; Table~\ref{tab:res}), when the mean cluster mass is about $5.0\times 10^{14} h^{-1}\,M_\odot$; by $z\approx 0.42$ (the mean redshift of our lowest redshift bin), $M_{\rm star,tot}=(2.5\pm 0.1)\times 10^{12}\,h^{-2}\,M_\odot$ while the clusters have grown to $M_{200m}=6.5\times 10^{14} h^{-1}\,M_\odot$.
Over the 4.3\,Gyr time span between $z=0.99$ and 0.3, 
we find that, for all three cluster mass bins, both total cluster mass and stellar mass content in red galaxies grow in a similar fashion (by around 30\%); specifically, the increase in $M_{\rm star,tot}$ is 32\%, 32\%, and 35\% for the low, middle, and high mass bins, suggesting a very weak, if at all, dependence on host cluster mass.  An implication of this result is that $M_{\rm star,tot}$ can be regarded as a good cluster mass proxy (e.g., \citealt{huang20}).

We next examine the stellar mass assembly history of BCGs (Figure~\ref{fig:newb}).
It is clear that the higher the cluster mass, the higher the BCG stellar mass (when comparing  the 3 sets of clusters at a given redshift), which is consistent with the expectation of the behavior of the (central) stellar mass--halo mass relation at the massive end (e.g., \citealt{lin04b,kravtsov18,wechsler18}).
We also find that, for the low and middle cluster mass bins, there is about 17\% and 12\% growth from $z=0.99$ to $z=0.3$, while there is not much growth for BCGs in the high mass bin.  This is reminiscent of the ``downsizing'' phenomenon, in the sense that BCGs in more massive clusters appear to have assembled (most of) their stellar mass earlier than their counterparts in lower mass clusters (at least in the central parts of the BCGs).
Finally, we note that for all 3 mass bins, there is very little growth at $z\lesssim 0.6$, a result consistent with previous findings \citep[e.g.,][]{lin13,lin17,bellstedt16,cerulo19}.  We caution, however, that we have simply used the {\tt cmodel} photometry for the galaxies throughout this analysis.  To truly capture the ``total'' luminosity of giant galaxies like BCGs, sophisticated analysis such as that of \citet{huang18} should be employed (see also \citealt{hsu22}).

\begin{figure}
    \plotone{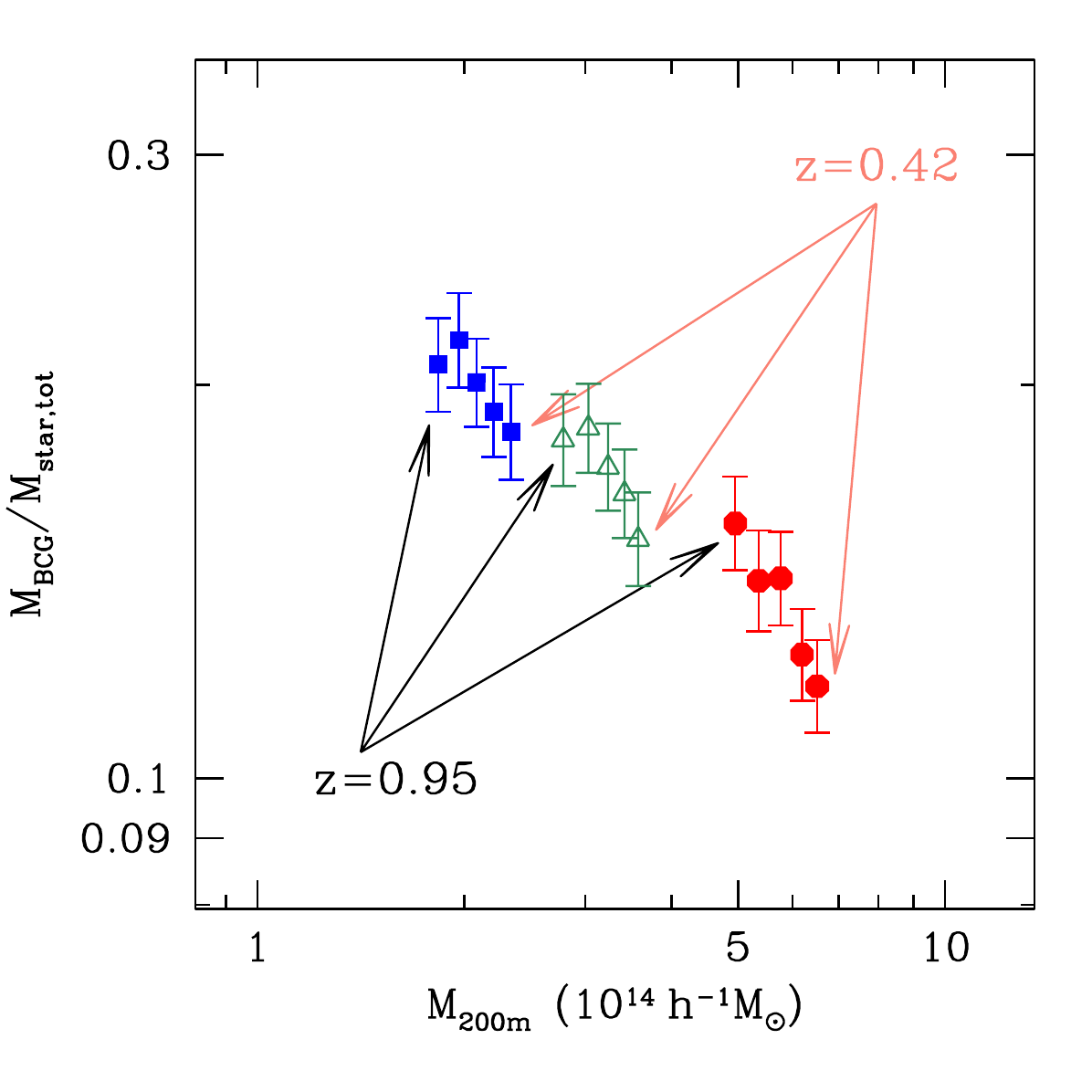}
    \caption{ Similar to Figure~\ref{fig:newm}, but for BCG-to-total stellar mass ratio. For a given cluster mass bin, the ratio decreases with time, because of the dramatic differences in the rates of growth between $M_{\rm bcg}$ and $M_{\rm star,tot}$.
    Also, at a given redshift, the BCG dominance decreases with cluster mass.
     }
    \label{fig:bm} 
\end{figure}

Finally, we present the BCG-to-total stellar mass ratio in Figure~\ref{fig:bm}.  Given the minute growth in BCG stellar mass compared to that of $M_{\rm star,tot}$ between $z=0.99$ and $z=0.3$, it is not surprising to see the steep decreasing trend with redshift, for all 3 cluster mass bins.
Furthermore, at a given epoch, the BCGs become less dominant as cluster mass increases (for example, compare the 3 points pointed by the black arrows or those three pointed by the pale red arrows).  This is consistent with the earlier findings for $z\approx 0$ clusters \citep[e.g.,][]{lin04b,gonzalez13}, as well as that from the XXL groups and clusters that span the redshift range of $z=0-1$ \citep{akino22}.
Comparing the typical value of $M_{\rm bcg}/M_{\rm star,tot} \approx 0.1$ inferred from the $K_s$-band luminosity (a good stellar mass proxy) for the $z<0.1$ cluster sample of \citet{lin04b}, the ratios shown in Figure~\ref{fig:bm} are clearly higher.  This is expected, as long as the trends shown in Figures~\ref{fig:newm} and \ref{fig:newb} continue to hold to lower-$z$ -- that is, the ratio will keep dropping when there is  a halt in BCG mass growth, while the total stellar mass continues to increase with time.

\subsection{Evolution of Scaling  Relations}
\label{sec:evsc}

\begin{figure}
    \epsscale{1.1}
    \plotone{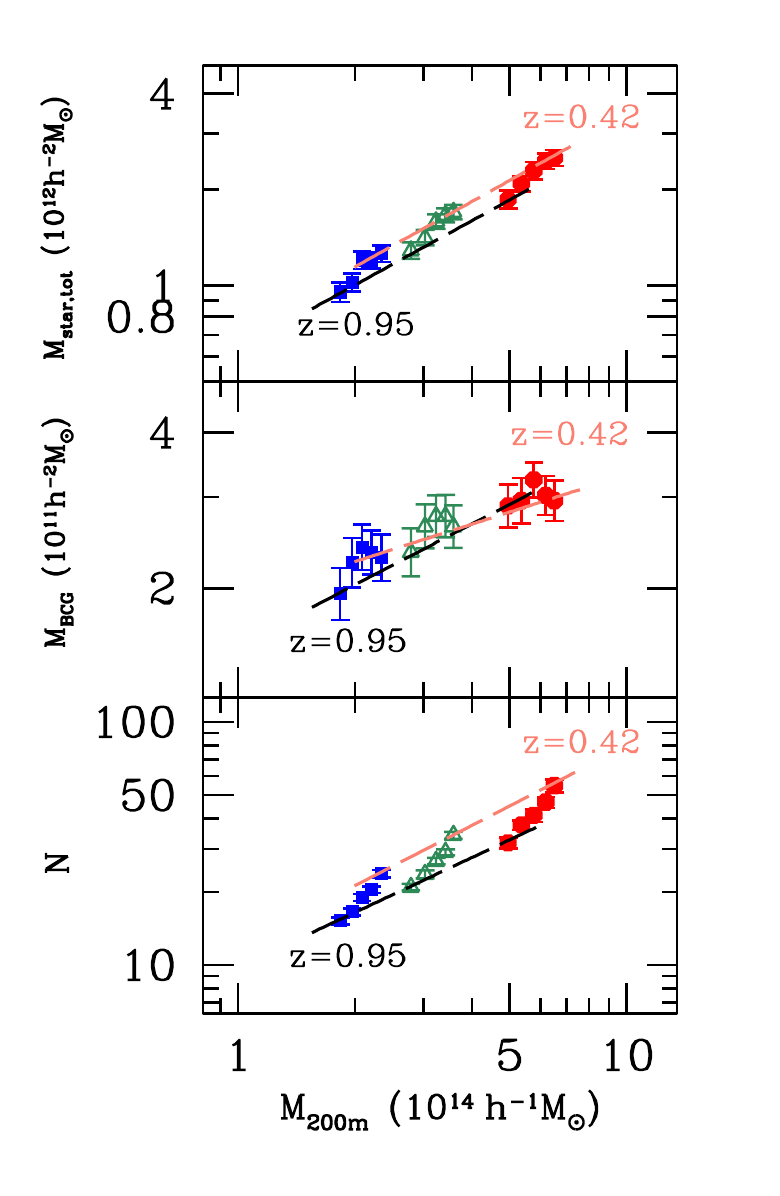}
    \caption{ Illustration of how our method can be used to derive mass--observable relations at different redshifts.  From top to bottom, we show example best fits to the $M_{\rm star,tot}$--$M_{200m}$, $M_{\rm bcg}$--$M_{200m}$, and $N$--$M_{200m}$ relations at $z\approx 0.95$ (black dashed line) and $z\approx 0.42$ (pale red dashed line).  For the values of the slopes, please refer to Table~\ref{tab:res}.  Here we simply note that the slopes of both $M_{\rm star,tot}$--$M_{200m}$ and $N$--$M_{200m}$ appear to be invariant with respect to redshift. 
     }
    \label{fig:3p} 
\end{figure}

In addition to being able to ``follow'' the redshift evolution for clusters with a specified low-$z$ mass, our method also enables a new way to examine how various mass--observable scaling relations (e.g., $M_{\rm star,tot}$--$M_{200m}$) evolves with time.
Given the total stellar mass of the three mass bins for each redshift, we use the least square fits to find the slope $\alpha$ as a function of redshift (assuming $M_{\rm star,tot} \propto M_{200m}^\alpha$).  
Two example fits are shown in the top panel of Figure~\ref{fig:3p} (the black and pale red dashed lines represent the best fits, with $\alpha=0.67\pm 0.09$ at $z\approx 0.95$ and $\alpha=0.68\pm 0.08$ at $z\approx 0.42$).
It is found that $\alpha$ scatters around 0.68 at all redshifts (see column 6 of Table~\ref{tab:res}).

In \citet{lin17} we traced the evolution of $M_{\rm star,tot}$ between $z\approx 1$ and $0.3$, for clusters whose mass at $z=0.3$ is quite close to our middle mass bin.  While we included both blue and red galaxies when computing $M_{\rm star,tot}$ in \citet{lin17} and found that the stellar mass grows as $M_{\rm star,tot} \propto M_{200c}^{0.7}$, if we only consider red galaxies, it is found that the growth steepens a bit to $M_{\rm star,tot} \propto M_{200c}^{0.8}$, which, given the uncertainties in $\alpha$ shown in Table~\ref{tab:res}, is  consistent with our result here.
Furthermore, our result supports the notion that the scaling relation shows very little change in slope, which has been established by several studies \citep[e.g.,][]{lin04,lin06,lin12,chiu16b,hennig17,chiu18}.  
The approach presented here  provides a fresh way to examine the way cluster galaxies evolve with time.
As to why the slope does not change with time, it remains an unsolved issue in cluster evolution (likely related to the formation of instracluster light; e.g., \citealt{golden-marx22}).  We refer interested readers to \citet[][Section 3.2 therein]{lin17} for an in-depth discussion.

In the middle panel of Figure~\ref{fig:3p}, we show example fits to the $M_{\rm bcg}$--$M_{200m}$ relation (again in the form $M_{\rm bcg} \propto M_{200m}^\beta$).  Although the slope at $z\approx 0.95$ is a bit high ($\beta=0.39\pm 0.15$), for the other 4 redshift bins, $\beta$ scatters between 0.2 and 0.3 (Table~\ref{tab:res}, column 7), which is consistent with results of \citet{lin04} and \citet{zhang16}, but appears to be shallower than other studies \citep[e.g.,][]{gonzalez13,chiu16,kravtsov18,golden-marx18,akino22}.  As the BCG stellar mass depends critically on how the photometry is measured (e.g., depth and aperture; \citealt{bernardi13}), it is perhaps not surprising to see a large range of $\beta$ in the literature \citep[e.g.,][]{golden-marx23}.  Using the central stellar velocity dispersion of BCGs as a mass proxy may be a promising alternative \citep{sohn22b}.

We show the $N$--$M_{200m}$ relation ($N \propto M_{200m}^\gamma$) and two example fits in the bottom panel of Figure~\ref{fig:3p}.  The best fit slope $\gamma$ scatters around 0.79 (Table~\ref{tab:res}, column 8).  The value found here is consistent with results in the literature mentioned above, and does not show much dependence on redshift (e.g., \citealt{lin06,costanzi21}; but see \citealt{capasso19} for a different result).  
We caution that, since we infer the halo mass from richness via the PDF derived from the $N$-$M_{200m}$ relation, the scaling relations shown here should be taken with a grain of salt and not be over-interpreted.

Finally, we recall that we have split the {\sc camira} clusters into two  sets (Section~\ref{sec:milsim}). Using only one of the two sets of the clusters does not change the results and conclusions presented here and in Section~\ref{sec:evol}, as for the fits to the scaling relations shown in the last 3 columns of Table~\ref{tab:res}, the values differ at only $\sim 5\%$ level.

\section{Discussion} 
\label{sec:disc}

We have presented a new way to study the evolution of clusters, which is in spirit similar to halo abundance matching, yet we further make use of dark matter halos trees to ``connect'' clusters at different cosmic epochs (c.f.~\citealt{behroozi13c}).
This method naturally incorporates the hierarchical structure buildup (and its stochastic nature), while simultaneously takes into account the uncertainty in estimation of halo mass via richness.
The method enables us to examine the time evolution of total stellar mass content in red cluster member galaxies ($M_{\rm star,tot}$) and the stellar mass assembly history of BCGs ($M_{\rm bcg}$), for 3 cluster subsamples that are chosen to lie in 3 different total mass ranges at $z=0.3$, the lower redshift limit of our study.
Furthermore, we are able to measure the redshift evolution of various mass--observable scaling relations, finding good agreement with previous studies, which corroborates the validity of our novel approach.

Our method is rather general, and can be applied to, for example, the stellar mass distribution (of both red and blue galaxies), in which case one has to keep track of the galaxy number counts as a function of stellar mass both within the cluster region and in a control (background) field.  Another possibility is to trace the abundance of certain population of galaxies (in the current paper, the red members), such as star-forming ones, or various types of active galactic nuclei.  
Now that the early eROSITA cluster sample and its mass--X-ray observable scaling relations are available (e.g., \citealt{bulbul24,grandis24}), it is possible to study how the mass of intracluster medium, or more generally the baryon content, evolves with redshift, for clusters of different final masses, which could provide strong constraints on cluster formation models.

In this work, due to practical reasons (Section~\ref{sec:over}), we limit ourselves to clusters at $z\le 1$.
In the near future, when the joint dataset  from {\it Euclid} \citep{laureijs11,euclid22b} and {\it Legacy Survey of Space and Time} \citep{ivezic19} becomes available, it will be possible to study cluster evolution all the way to $z\sim 2$.

An ultimate validation of our method would be applying it to a full hydrodynamical or semi-analytic simulation (which requires a full modeling of a mass--observable relation, itself a non-trivial task), and see if the evolution of galaxies or baryonic components can be recovered.  As a work of exploratory  nature, we here focus on advocating the power of this approach, and will leave the full validation, along with a detailed investigation of cluster galaxy evolution (such as the stellar mass distribution as mentioned above) to a future work.


\begin{acknowledgments}
We thank an anonymous referee for comments that improved the clarity of this paper.
We thank Ryoma Murata for helpful comments on the inversion of the mass--richness scaling relation, and Gerald Lemson for the help in  querying  the Millennium database efficiently.
We  appreciate helpful comments from Christian Kragh Jerspersen, Ting-Wen Lan, Chun-Hao To, and Jubee Sohn.
We are grateful for support from the National Science and Technology Council  of Taiwan under grants MOST 111-2112-M-001-043, NSTC 112-2112-M-001-061, and NSTC 113-2112-M-001-005.
This work was supported by JSPS KAKENHI Grant Numbers JP23K22531, JP20H05856, JP24K00684.
YTL thanks Institute for Advanced Study for its hospitality during the final stage of completion of this work, and is grateful to the Bershadsky Fund for support.
YTL thanks IH, LYL and ALL for constant encouragement and inspiration.

The Hyper Suprime-Cam Subaru Strategic Program (HSC-SSP) is led by the astronomical communities of Japan and Taiwan, and Princeton University.  The instrumentation and software were developed by the National Astronomical Observatory of Japan (NAOJ), the Kavli Institute for the Physics and Mathematics of the Universe (Kavli IPMU), the University of Tokyo, the High Energy Accelerator Research Organization (KEK), the Academia Sinica Institute for Astronomy and Astrophysics in Taiwan (ASIAA), and Princeton University.  The survey was made possible by funding contributed by the Ministry of Education, Culture, Sports, Science and Technology (MEXT), the Japan Society for the Promotion of Science (JSPS),  Japan Science and Technology Agency (JST),  the Toray Science Foundation, NAOJ, Kavli IPMU, KEK, ASIAA,  and Princeton University.

The Millennium simulation databases used in this paper and the Web application providing online access to them were constructed as part of the activities of the German Astrophysical Virtual Observatory (GAVO).
\end{acknowledgments}

\end{document}